\documentclass[12pt]{article}
\usepackage{epsfig}
\usepackage{float}
\usepackage{graphicx}
\usepackage{amsmath}
\usepackage{amssymb}
\usepackage[noadjust]{cite}

\begin{document}
\bigskip

\newcommand{\be}{\begin{equation}}
\newcommand{\ee}{\end{equation}}
\newcommand{\noi}{\noindent}
\newcommand{\ra}{\rightarrow}
\newcommand{\bib}{\bibitem}
\newcommand{\refb}[1]{(\ref{#1})}
\newcommand{\bff}{\begin{figure}}
\newcommand{\eff}{\end{figure}}


\begin{center}
{\Large\bf Extreme   Dilaton Black Holes in 2 +1 Dimensions: Quasinormal modes}

\end{center}
\hspace{0.4cm}
\begin{center}

Sharmanthie Fernando \footnote{fernando@nku.edu}\\
{\small\it Department of Physics \& Geology, Northern Kentucky University, Highland Heights}\\
{\small\it Kentucky 41099, U.S.A.}\\

P.~A.~Gonz\'{a}lez \footnote{pablo.gonzalez@udp.cl}\\
{\small \it Facultad de Ingenier\'{\i}a y Ciencias, Universidad Diego Portales, Avenida Ej\'{e}%
rcito Libertador 441, Casilla 298-V, Santiago, Chile.}\\

Yerko V\'{a}squez \footnote{yvasquez@userena.cl}\\
{\small \it Departamento de F\'{\i}sica, Facultad de Ciencias, Universidad de La Serena,\\ 
Avenida Cisternas 1200, La Serena, Chile.}

\end{center}


\begin{center}
{\bf Abstract}
\end{center}

\hspace{0.7cm} 

We study neutral massless scalar field perturbations around an extreme  dilaton black hole in 2 +1 dimensions: the wave equations of the massless  scalar field is shown to be exactly solvable in terms of Whittaker  functions. Thus, the quasinormal modes are computed exactly and shown to be   purely imaginary: we show the existence of stable and unstable modes. Interestingly,  the quasinormal modes do not depend on the black holes parameters and the fundamental mode is always unstable and depends only on the parameters of the test field.  Also, we determine the quasinormal frequencies via the improved asymptotic iteration method which shows a good agreement with the analytical results. \\


\clearpage
\tableofcontents


\clearpage

\section{Introduction}

Extreme black holes have special geometry since their inner horizon and outer horizon coincide. Also, surface  gravity of such black holes is zero, leading to zero temperature. Furthermore, semi classically it has been shown that some extreme black holes have zero entropy (even though the area is non-zero) \cite{ross1}. Extreme black holes are important objects due to many reasons. They are considered important in supergravity theories, string theory and M theory. For example, extreme Reissner-Nordstrom black hole can be embedded in N= 2 supergravity theory \cite{kallosh1}. Some supersymmetric black holes (such as the extreme Reissner-Nordstrom black hole) do have extra symmetry that  non-extreme black holes don’t have: they admit Killing spinor fields leading to the extreme black hole to be invariant  under supersymmetric transformations. Extreme black holes take a special place in counting micro states related to entropy in black holes: large class of extremal supersymmetric black holes in string theory have been studied in the counting of mircrostates \cite{sen1,sen2}.\\

The Quasinormal Modes (QNMs) and  Quasinormal Frequencies (QNFs) in black hole backgrounds has been studied for many years  starting from the pioneering work by Regge and Wheeler \cite{Regge:1957a,Zerilli:1971wd,
Zerilli:1970se, Kokkotas:1999bd, Nollert:1999ji, Konoplya:2011qq}. Furthermore, recent detection of gravitational waves \cite{Abbott:2016blz, TheLIGOScientific:2016src} have stimulated further the study of such vibrations since QNM   emerge as the final stage of large coalescing objects. \\

While there are many works on QNM’s of  non-extreme black holes, there are only few works on extreme black holes on the topic. The extreme rotating BTZ black hole was studied in \cite{Crisostomo:2004hj} and was shown the absence of QNM frequencies for scale and spiner fields. In 4 dimensions, neutral scalar field perturbations around the Reissner-Nordstrom black hole was studies in \cite{Onozawa:1995vu}. Charged massless perturbations around an extremal Reissner-Nordström black hole and to neutral massless perturbations around an extremal Kerr black hole was studied in Ref. \cite{Richartz:2015saa}, and it was shown that the QNMs spectrum presents a decay rate. Also, there are studies about QNMs at the near extremal limit, where in general it was shown that in this limit the near extremal modes are dominant, and for uncharged scalar fields these modes become purely imaginary, see \cite{Berti:2003ud, Richartz:2014jla, Cardoso:2017soq,  Cardoso:2018nvb, Destounis:2018qnb, Panotopoulos:2019tyg, Liu:2019lon, Destounis:2019hca, Destounis:2019omd, Destounis:2020pjk, Destounis:2020yav, Aragon:2021ogo, Fontana:2020syy}. \\

The three-dimensional models of gravity have been of great interest due to their simplicity over the four-dimensional and higher-dimensional models, and since some of their properties are shared by their higher-dimensional analogs. The black hole considered in this paper is a solution to the Einstein-Maxwell-dilaton gravity in 2 + 1 dimensions. The corresponding action is given as follows,
\begin{equation} \label{action}
S = \int d^3x \sqrt{-g} \left[ R - 4  (\bigtriangledown \phi )^2 -
e^{-4  \phi} F_{\mu \nu} F^{\mu \nu} + 2 e^{4 \phi} \Lambda \right]\,.
\end{equation}
Here, $R$ is the scalar curvature, $ \Lambda$ is treated as the cosmological constant, $\phi$ is the dilaton field and $F_{\mu \nu}$ is the Maxwell's field strength. The action in Eq. $\refb{action}$ is conformally related to the low-energy string action in 2+1 dimensions. The dilaton field plays the role of the extra fields, which naturally arises, for instance, in the compactifications from higher-dimensional models or from string theory. These theories also have black hole solutions which play an important role in revealing various aspects about the geometry of spacetime and the quantization of gravity, and also the physics related to string theory \cite{Witten:1991au, Teo:1998kp,McGuigan:1991qp}. \\

In this paper, we will study the extreme dilaton black holes in 2 +1 dimensions for the above theory. In particular we will study the propagation of neutral massless scalar fields around the black hole. Some works related to QNM's for dilaton black holes are given in  \cite{Ferrari:2000ep, Konoplya:2001ji, Fernando:2003ai, Kettner:2004aw, Chen:2004zr,Chen:2005rm, Chen:2005pv, Fernando:2008hb, Myung:2008pr, Lin:2010zzf, Sakalli:2013yha, Becar:2014jia, Fernando:2016ftj, Blazquez-Salcedo:2017txk,Destounis:2018utr, Brito:2018hjh, Zinhailo:2019rwd, Rincon:2020pne}. For the spacetime under study the QNM spectrum is characterized by purely imaginary modes and we will show that the fundamental mode depends only on the properties of the test field. Also, we discuss about the stability of the propagation of neutral massless scalar fields in this extremal background. In addition, we apply the improved Asympotic Iteration Method (AIM) to evaluate the QNFs numerically and we discuss about their accuracy.\\

This work is organized as follows. In Sec.~\ref{background} we give a brief review of three-dimensional extreme dilaton black hole. Then, in Sec.~\ref{perturbations} we study neutral massless scalar perturbations and we present an exact solution to the wave equation in Sec.~\ref{exact}, and to the QNMs in Sec.~\ref{QNM}. Then, in section ~\ref{numerically} we study the spectrum  numerically by using the improved AIM method. Finally, we conclude in Sec.~\ref{conclusion}.


\section{Extreme  dilaton black hole}
\label{background}

In \cite{chan2}, Chan and Mann derived  static changed solutions to the above action  in Eq. $\refb{action}$ as,
$$
ds^2= - f(r)dt^2 +  \frac{4 r^2 dr^2}{f(r)} + r^2 d \theta^2\,,
$$
\begin{equation} \label{metric}
f(r) =\left( -2Mr + 8 \Lambda r^2 + 8 Q^2 \right); \hspace{0.1cm} \phi = \frac{1}{4}  ln (\frac{r}{\beta}) ; \hspace{1.0cm}F_{rt} = \frac{Q}{r^2}\,.
\end{equation}
As was discussed in \cite{chan2}, the above solutions represents a black hole for $M \geq 8 Q \sqrt{\Lambda}$ and $\Lambda >0$. For $M > 8 Q \sqrt{\Lambda}$, the space-time has two horizons given by the zeros of $g_{tt}$;
\begin{equation}
r_+ =  \frac{M + \sqrt{ M^2 - 64 Q^2 \Lambda}}{8 \Lambda}; \hspace{1.0cm}
r_- = \frac{M - \sqrt{ M^2 - 64 Q^2 \Lambda}}{8 \Lambda}\,.
\end{equation}
The given  black hole is also a solution to low energy string action  by a  conformal transformation,
\begin{equation}
g^{String} =  e^{4 \phi} g^{Einstein}\,.
\end{equation}

In this paper, we will focus on the extreme changed black hole where $M= 8 Q \sqrt{\Lambda}$. Then the function $f(r)$ becomes,
\begin{equation}
f(r) = 8 \Lambda ( r - r_h)^2\,,
\end{equation}
and the black hole has only one horizon given by,
\begin{equation}
r_h = \frac{M}{8 \Lambda}\,.
\end{equation}
There is a time-like singularity at $r =0$. As explained in \cite{chan2}, the black hole space-time in neither de-Sitter ($\Lambda < 0$) nor anti-de-Sitter ($\Lambda > 0$). The Hawking temperature  $T_H =0$.


\section{Neutral scalar perturbation of  the extreme dilaton  black hole}
\label{perturbations}

In this section, we will develop the equations for a neutral scalar field in the background of the extreme charged dilaton black hole. The equation is similar to what was presented for the non-extreme black hole in \cite{fer2}. 

\begin{figure}
\begin{center}
\includegraphics[width=0.47\textwidth]{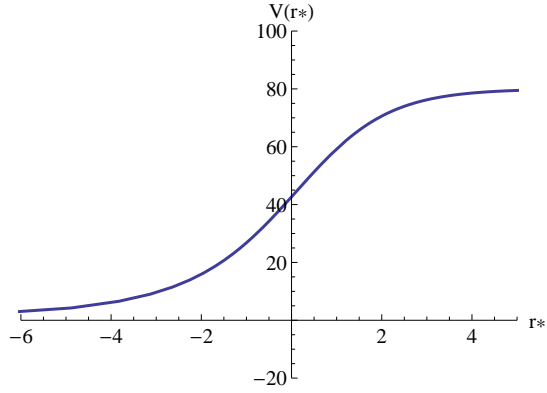}
\end{center}
\caption{The behavior of the potential $V(r)$ plotted against $r_*$. Here,  $M = 10, m= 2$ and  $\Lambda = 2$.}
\label{potential1}
\end{figure}

The equation for a massless neutral scalar field in curved space-time can be written as,
\begin{equation} \label{klein}
\bigtriangledown ^{\mu} \bigtriangledown_{\mu}  \Phi     =  0\,.
\end{equation}
Using the following anzatz,
\begin{equation} 
\Phi = e^{ i m \theta} e^{ - i \omega t }\frac{\xi(r)} { \sqrt{r}}
\end{equation}
Eq. $\refb{klein}$  simplifies into  the equation
\begin{equation}
\left(\frac{d^2 }{dr_*^2} + \omega^2    - V(r) \right) \xi(r_*) = 0\,.
\end{equation}
Here, $V(r)$ is given by,
\begin{equation}
V(r) =  \frac{f(r) } { 2 r^{3/2} } \frac{d}{dr} \left( \frac{ f(r) } { 4 r^{3/2} } \right) + \frac{ m^2 f(r)}{r^2}\,. 
\end{equation}
When expanded, $V(r)$ is given as,
\be
V(r) =  8 m^2 \Lambda + 4 \Lambda^2  - \frac{ 2 m^2 M}{ r}  + \left( \frac{- 3 M^2}{8} + \frac{ m^2 M^2}{ 8 \Lambda} \right) \frac{1}{r^2}   + \frac{ M^3}{ 16 \Lambda r^3} - \frac{ 3 M^4}{ 1024 \Lambda^2 r^4}\,.
\ee
Also, $r_{*}$ is the tortoise coordinate computed as,
\begin{equation} \label{tor}
dr_{*} = \frac{2 r dr}{f(r)} \Rightarrow 
r_* = \frac{ 1}{ 4 \Lambda } \left(  ln( r - r_h) - \frac{r_h}{ r - r_h} \right)\,.
\end{equation}
Note that when $r \rightarrow r_h$ , $r_* \rightarrow - \infty$ and for $r \rightarrow \infty$, $ r_* \rightarrow \infty $. In Fig. \ref{potential1}, the potential $V(r)$ is plotted as a function of $r_*$. Now, a discussion about the behavior of the potential is in order. Notice that the potential behaves like a step function with it approaching a constant value. In fact the constant value is $ V_0 = 8 m^2 \Lambda + 4 \Lambda^2$. When a potential behaves as a step function, the quasinormal modes are pure imaginary. This was discussed in length  in \cite{yun}. Another example where pure imaginary quasinormal modes arise with such a step function is  the five dimensional dilaton black hole \cite{lopez}. Hence just by observing the from of the potential, one can predict that the quasinormal modes for the extreme black hole will be pure imaginary. In the next section, we will solve the equations exactly and prove that is indeed the case.

\newpage



\section{Exact solutions to the  scalar wave equation}
\label{exact}

In order to find exact solutions to the wave equation for the massless  scalar, we will revisit the Eq. $\refb{klein}$ in section 3 with the anzatz,
\begin{equation}
\Phi =  e^{- i \omega t} e^{i m \theta} R(r) 
\end{equation}
Eq. $\refb{klein}$ leads to the radial equation,
\begin{equation} \label{wave}
\frac{d}{dr} \left( \frac{f(r)}{2} \frac{dR(r)}{dr} \right) + 2r^2 \left( \frac{\omega^2}{f(r)}   - \frac{m^2}{r^2}  \right)  R(r)  = 0\,.
\end{equation}
In order to solve the wave equation exactly, we will redefine  
$r$ coordinate of the Eq. $\refb{metric}$ with a new variable $x$ given by,
\begin{equation}
 x = \left( \frac{ r_h}{ r - r_h} \right)\,.
\end{equation}
In the new coordinate system, $x = 0$ corresponds to $r \rightarrow \infty$ and $x = \infty$ corresponds  $r = r_h$. With the new coordinate,  Eq. $\refb{wave}$ becomes,
\begin{equation} \label{wave2}
 \frac{d^2 R}{dx^2} +  \left( \frac{A} {x^2} + \frac{ B} {x} + C \right)R =0
\end{equation}
where,
$$
A=  \frac{ \omega^2}{ 16 \Lambda^2} - \frac{ m^2}{ 2 \Lambda}\,,$$
$$
B = \frac{ \omega^2}{ 8 \Lambda^2}\,,$$
\begin{equation}
C =  \frac{\omega^2} {16  \Lambda^2}\,.
\end{equation}
The Eq. $\refb{wave2}$ resembles the Whittaker differential equation given by \cite{math},
\begin{equation}
 \frac{d^2 R}{dz^2} +  \left( \frac{\frac{1}{4} - \mu^2} {z^2} + \frac{ \kappa} {z}  - \frac{1}{4}\right)R =0\,.
\end{equation}
The solutions to the Whittaker equation are given by,
$$
M_{ \kappa, \mu} (z) = e^{-\frac{z}{2} } z^{ \frac{1}{2} + \mu} M[ \frac{1}{2} + \mu - \kappa , 1 + 2 \mu, z]\,,$$
and
\begin{equation}
M_{ \kappa, -\mu} (z) = e^{-\frac{z}{2} } z^{ \frac{1}{2} - \mu} M[ \frac{1}{2} - \mu - \kappa , 1 - 2 \mu, z]\,.
\end{equation}
The function $M[a,b,z]$ is called the $Kummer Function$ and is the solution to the Kummer differential equation given by \cite{math},
\begin{equation}
z\frac{d^2 F}{dz^2} + ( b -z) \frac{d F}{dz} -a F =0\,.
\end{equation}
Now, to transform the Eq. $\refb{wave2}$ to the Whittaker form, we will reparametrize the coordinate $x$ as
\begin{equation}
y = i 2 \sqrt{C} x = \frac{i \omega} { 2 \Lambda} x\,.
\end{equation}
Then the Eq. $\refb{wave2}$ takes the form,
\begin{equation}
 \frac{d^2 R}{dy^2} +  \left( \frac{\frac{1}{4} - \mu^2} {y^2} + \frac{ \kappa} {y}  - \frac{1}{4}\right)R =0\,,
\end{equation}
with
\begin{equation}
\kappa = \frac{ i \omega}{ 4 \Lambda}; \hspace{0.5cm} \mu = \frac{i}{2}  \sqrt{\frac{  \omega^2 - 8 m^2 \Lambda}{4 \Lambda^2} - 1 }\,.
\end{equation}
Note that we will take $``+''$ sign without loss of generality. Before presenting the final form of the solution to the wave equation, we will redefine the parameters $\kappa$ and $\mu$ with $\alpha$ and $\beta$ as,
\begin{equation} \label{newdef}
\kappa = \alpha; \hspace{2 cm} \mu = \beta - \frac{1}{2}\,.
\end{equation}
The reason to introduce the new parameters is to follow the same notation followed with the computation done for the non-extreme black hole by the same author in \cite{fer2}. Finally, by substituting the values of $\kappa$ and $\mu$ in terms of $\alpha$ and $\beta$, the general solution to the wave equation can be written as
\begin{equation}
R(y) = C_1   e^{-\frac{y}{2} } y^{ \beta} F[ \beta - \alpha , 2 \beta, y] + C_2   e^{-\frac{y}{2} } y^{ 1 - \beta} F[1- \beta - \alpha, 2 ( 1 - \beta), y]\,.
\end{equation}
Hence the general solution to the waves equation is,
\begin{equation}
\Phi( y, t, \theta) =  e^{  i m \theta} e^ { - i \omega t} R(y)\,.
\end{equation}


\section{Quasinormal modes of the extreme dilaton black holes}
\label{QNM}

Since the main objective of the paper is to find QNMs of the extreme black hole, one needs to impose specific boundary conditions to the general solution obtained in section 4. The solutions are analyzed closer to the horizon and at infinity to obtain exact results for QNMs.

To compute QNMs of a black hole, two boundary conditions are imposed. One is to impose the wave to be purely ingoing at he horizon. The other is the boundary  condition for large $r$ values. For asymptotically flat space-times, the asymptotic boundary condition is for the field to be purely out going. For non-asymptotically-flat space-times, there are two possibilities: on is for the field to vanish and the other is for the flux to vanish. We choose the field to vanish similar to what was done in \cite{fer2}.


\subsection{Solution at asymptotic region}

First we will study what  the solution  is when $r \rightarrow \infty$. 
Since for large $r$, $ y \rightarrow 0$, the Kummer function $M[a,b,y] \rightarrow 1$. By substituting $ y \rightarrow 0$  to the exact solution obtained for $R(y)$, one obtain,
\begin{equation} \label{larger}
R(r \ra \infty, r \ra 0) \approx  C_1 y^{\beta} + C_2 y^{ 1 - \beta}\,.
\end{equation}
Since,
\be
y =  \frac{ i \omega r_h}{  2 \Lambda( r - r_h)}
\ee
for large $r$,
\begin{equation}
 y \rightarrow \frac{ i \hat{\omega}}{ 2 \Lambda} \left( \frac{r_h}{r} \right)
\end{equation}
the Eq. (\ref{larger}) can be written in terms of $r$ as,
\begin{equation} \label{larger2}
R(r \rightarrow \infty) \approx  C_1 \left( \frac{ i \hat{\omega}}{ 2 \Lambda} \right)^{\beta}\left( \frac{r_h}{r} \right)^{\beta} + C_2 \left( \frac{ i \hat{\omega}}{ 2 \Lambda} \right)^{1 -\beta} \left( \frac{r_h}{r} \right)^{1 - \beta}\,.
\end{equation}
Now one need to  determine which part of the solution in Eq. (\ref{larger2}) corresponds to the ``ingoing'' and ``outgoing'' respectively. For that  we will first find the tortoise coordinate $r_{*}$ in terms of $r$ at large $r$. Note that for large $r$, $f(r) \rightarrow 8 \Lambda r^2 $. Hence  the equation relating the tortoise coordinate $r_*$ and $r$ in Eq. $\refb{tor}$ simplifies to,
\begin{equation}
dr_{*} = \frac{ dr}{ 4 \Lambda r}\,.
\end{equation}
The above can be integrated to obtain,
\begin{equation}
r_{*} \approx  \frac{1}{4 \Lambda} ln( \frac{ r}{r_h} )\,.
\end{equation}
Hence,
\begin{equation} \label{newtor}
r \approx r_h e^{ 4 \Lambda r_*}\,.
\end{equation}
Substituting $r$ from Eq. (\ref{newtor}) and $\beta$ from Eq. (\ref{newdef}) into the Eq. (\ref{larger2}), $R(r \rightarrow \infty)$ is rewritten as,
$$
R(r  \rightarrow \infty ) \rightarrow C_1 \left( \frac{ i \hat{\omega}}{ 2 \Lambda} \right)^{\beta} e ^{ -i  \omega r_* \sqrt{1 - \frac{ 4 \Lambda^2}{ \omega^2} ( \frac{2 m^2}{\Lambda} +1) } - 2 \Lambda r_*} $$
\begin{equation}
+ C_2 \left( \frac{ i \hat{\omega}}{ 2 \Lambda} \right)^{1 -\beta} e ^{ i  \omega r_* \sqrt{1 - \frac{ 4 \Lambda^2}{ \omega^2} ( \frac{2 m^2}{\Lambda} +1) } - 2 \Lambda r_*}\,. 
\end{equation}
From the above one can conclude  that the first term and the second term represents the ingoing and outgoing waves respectively. Since for QNMs, the ingoing amplitude has to be zero, we chose $C_1 =0$. Hence the solution to the wave equation becomes,
\begin{equation}
R(y) =  C_2   e^{-\frac{y}{2} } y^{ 1 - \beta} M[1- \beta - \alpha, 2 ( 1 - \beta), y]\,.
\end{equation}


\subsection{Solutions near the horizon}

Now we can analyze the solution of the wave equation near the horizon ($r \rightarrow r_h$, 
$y \rightarrow \infty$). The Kummer function $M[a,c,y]$ has an expansion for large $y$ as follows \cite{math},
\begin{equation}
F[a,c,y] \rightarrow e^{y} y^{ a -c} \frac{ \Gamma( c)}{\Gamma( a)}  +  e^{ \pm i \pi a } y^{- a} \frac{ \Gamma(c) }{ \Gamma( c -a)}\,.
\end{equation}
Note that the upper sign is taken if $ - \frac{\pi} {2} < arg(y) < \frac{3 \pi} {2}$ and lower sign is chosen $ - \frac{3 \pi}{2} < arg(y) < - \frac{ \pi}{2}$. Since   $arg(y) = \frac{\pi}{2}$ the upper sign will be chosen. By substituting the above expansion with the appropriate values for $a$ and $c$ in terms of $\alpha$ and $\beta$, one obtain the function $R(y)$ 
near the horizon as,
\be
R( r \rightarrow r_h, y \rightarrow \infty ) \rightarrow C_2 \left( e^{y/2} y^{-\alpha} \frac{ \Gamma( 2 - 2 \beta)}{\Gamma( 1 - \beta - \alpha)} + e^{ - y/2 + i \pi ( 1 - \beta - \alpha) } y^{ \alpha} \frac{ \Gamma(2 - 2 \beta)}{ \Gamma( 1 - \beta + \alpha)} \right)\,.
\end{equation}
To determine which part of the above equation represents the ingoing/outgoing, one has to introduce the tortoise coordinate near the horizon. From Eq. $\refb{tor}$, the tortoise coordinate near the horizon can be approximated with,
\begin{equation} \label{newtor2}
r_* \approx \frac{1}{ 4 \Lambda} ln( r - r_h) \rightarrow e^{ - 4 \Lambda r_*} \approx \frac{ 1}{ ( r - r_h)}\,.
\end{equation}
By substituting $y$ in terms of $r_*$ in Eq. $\refb{newtor2}$, the radial function $R(r_*)$ approximates to,
$$
R( r \rightarrow r_h, y \rightarrow \infty ) \approx C_2  e^{\frac{y}{2}}  \frac{ \Gamma( 2 - 2 \beta)}{\Gamma( 1 - \beta - \alpha)} \left(\frac{ i \hat{\omega} r_h}{ 2 \Lambda} \right)^{-\alpha} e ^{ i  \omega r_*}+$$
\begin{equation}
C_2 e^{ - \frac{y}{2} + i \pi ( 1 - \beta - \alpha) }  \frac{ \Gamma(2 - 2 \beta)}{ \Gamma( 1 - \beta + \alpha)} \left(\frac{ i \hat{\omega} r_h}{ 2 \Lambda} \right)^{\alpha} e ^{ - i  \omega r_*}\,.
\end{equation}
The first represents the outgoing waves and the second represents the ingoing wave at the horizon.


\subsection{Quasinormal modes}
Since the QNMs are defined with as purely ingoing waves at the horizon, the outgoing terms has to be zero. Since $C_2 \neq 0$, the first term vanish only at the poles of the Gamma function, $\Gamma(1 - \beta - \alpha)$. Note that the Gamma function $\Gamma(x)$ has poles at $ x = - n$ for $ n = 0,1,2..$. Hence to obtain QNMs, the following relations has to hold.
\begin{equation} \label{qnm1}
1- \alpha - \beta  = - n
\end{equation} 
leading to,
\begin{equation} \label{qnm2}
\beta = ( 1 + n) - \frac{ i \omega } { 4 \Lambda }\,.
\end{equation}
By combining Eq. $\refb{qnm1}$ and Eq. $\refb{qnm2}$, one can solve for $\omega$   as,
\begin{equation}
\label{omega}
\omega =  \frac{-2 i}{ 2n +1} \left( 2 \Lambda n (1+n) - m^2 \right)\,.
\end{equation}
They are pure imaginary, do not depend on the black hole parameters, and the decay rate increases when the cosmological constant increases. Note that the fundamental mode $n=0$ is unstable, and the decay rate depends only on the angular number of the test field. Due to the minus sign in front, these oscillations will be damped leading to stable perturbations for $2 \Lambda n (1+n) > m^2$. However, for $2 \Lambda n (1+n) < m^2$, the oscillations would lead to unstable modes. 

In acoustic black holes, QNM's for scalar field shows similar properties \cite{joel}. There, 
\begin{equation}
     \omega = -\frac{i}{2} \frac{(n-1) (n+3) \tilde{a}}{n+1} \,.
     \end{equation}
For $n=0$, $\omega$ is positive and will lead to exponentially growing mode similar to the extreme dilaton black hole discussed above.


\section{Numerical analysis}
\label{numerically}

Some well known numerical methods to obtain QNM frequencies are: the Mashhoon method, Chandrasekhar-Detweiler method, WKB method, Frobenius method, continued fraction method, asymptotic iteration method (AIM) and improved AIM, pseudospectral Chebyshev method, among others.  In this section, we apply the improved AIM \cite{Cho:2009cj} method to compare with the analytical  results.  This is an improved version of the method proposed in Refs. \cite{Ciftci, Ciftci:2005xn} and it has been applied successfully in the context of QNMs for different black holes geometries; see for instance \cite{Cho:2011sf, Catalan:2013eza, Zhang:2015jda, Barakat:2006ki, Sybesma:2015oha, Chen:2015jga, Catalan:2014ama,  Gonzalez:2015gla, Becar:2015kpa, Becar:2015gca}.

The boundary conditions satisfied by the QNMs are given by
\begin{eqnarray}
\nonumber    \xi \sim e^{-i \omega r_{\ast}}\,\,\, && \text{as} \,\,\,  r_{\ast} \rightarrow -\infty \\
    \xi \sim e^{i \tilde{\omega} r_{\ast}}\,\,\,
\nonumber      &&  \text{as} \,\,\,   r_{\ast} \rightarrow \infty
\end{eqnarray}
where $\tilde{\omega}^2 = \omega^2 - 4 \Lambda^2 - 8 \Lambda m^2$. There are only ingoing waves at the horizon and outgoing waves at infinity. This can be transformed to

\begin{eqnarray}
\nonumber    R \sim r^{-\frac{1}{2}}(r-r_h)^{-\frac{i \omega}{4 \Lambda}} e^{\frac{i \omega}{4 \Lambda} \frac{r_h}{r-r_h}}\,\,\,  && \text{as}\,\,\, r \rightarrow r_h \\ 
\nonumber    R \sim  r^{-\frac{1}{2}+\frac{i \tilde{\omega}}{4 \Lambda}} \,\,\,   &&  \text{as} \,\,\, r \rightarrow \infty
\end{eqnarray}
In order to write a solution with this behavior at the boundaries, we define

\begin{equation} \label{RR}
    R(r) = r^{-\frac{1}{2}+ \frac{i ( \tilde{\omega}+ \omega)}{4 \Lambda}} (r-r_h)^{- \frac{i \omega}{4 \Lambda}} e^{\frac{i \omega r_h}{4 \Lambda (r-r_h)}} \chi(r)\,.
\end{equation}
Inserting this expression in Eq. (\ref{wave}) and performing the change of variable $z = 1- r_h/r$ we arrive at the following equation

\begin{equation} \label{ec}
    \chi''(z) = \lambda_0(z) \chi'(z) + s_0 (z) \chi \,,
\end{equation}
where the prime denotes derivative with respect to $z$ and
\begin{eqnarray}
\nonumber \lambda_0(z) &=& - \frac{1}{2(1-z)z^2 \Lambda}\left( - i \omega  + 4 \Lambda z + z^2 \left(i \omega - 6\Lambda + \sqrt{-\omega^2 + 4 \Lambda^2 + 8 \Lambda m^2} \right)  \right) \\
\nonumber s_0(z) &=& - \frac{1}{8(1-z)z^2 \Lambda^2} \Big( \omega^2 - i \omega \sqrt{-\omega^2 + 4 \Lambda^2 + 8 \Lambda m^2} -4 \Lambda m^2 + z (\omega^2 + 4 i \omega \Lambda -   \\
&& 4 \Lambda m^2 - 8 \Lambda^2 + 4 \Lambda \sqrt{-\omega^2+ 4 \Lambda^2 + 8 \Lambda m^2} - i \omega \sqrt{-\omega^2+ 4 \Lambda^2 + 8 \Lambda m^2} )   \Big) \,.
\end{eqnarray}

We solve numerically Eq. (\ref{ec}) using the improved AIM. This method is implemented as follows, first it is necessary to differentiate Eq. (\ref{ec}) $n$ times with respect to $z$, what yields
\begin{equation}
    \chi^{n+2} = \lambda_n(z) \chi' +s_n(z) \chi \,,
\end{equation}
where
\begin{eqnarray} \label{expansion}
\nonumber \lambda_n(z) &=& \lambda'_{n-1}(z)+s_{n-1}(z)+\lambda_0(z)\lambda_{n-1}(z)   \\
s_n(z) &=& s'_{n-1}(z) +s_0(z)\lambda_{n-1}(z)\,.
\end{eqnarray}
Then, the functions $\lambda_n$ and $s_n$ are Taylor expanded around some point $z_0$ ($0<z_0<1$) at which the improved AIM is performed
\begin{eqnarray}
\nonumber \lambda_n(z_0) &=& \sum_{i=0}^{\infty} c_n^i (z-z_0)^i \\
\nonumber s_n(z_0) &=& \sum_{i=0}^{\infty} d_n^i (z-z_0)^i \,,
\end{eqnarray}
$c_n^i$ and $d_n^i$ are the i$^{th}$ Taylor coefficients of $\lambda_n(z_0)$ and $s_n(z_0)$ respectively. Then, replacing these Taylor expansions in Eqs. (\ref{expansion}) the following set of recursion relations for the coefficients are obtained
\begin{eqnarray}
\nonumber c_n^i &=& (i+1)c_{n-1}^{i+1}+d_{n-1}^i +\sum_{k=0}^i c_0^k c_{n-1}^{i-k}  \\
\nonumber d_n^i  &=&  (i+1) d_{n-1}^{i+1} + \sum_{k=0}^i d_0^k c_{n-1}^{i-k} \,.
\end{eqnarray}
Next, imposing a termination to the number of iterations one arrives to the following quantization condition

\begin{equation} \label{ecuacion}
    d_n^0 c_{n-1}^0 - d_{n-1}^0 c_n^0 = 0 \,.
\end{equation}
We use a root-finding algorithm to determine numerically the QNFs from (\ref{ecuacion}). In Table \ref{Table1} we show some values of QNFs, in order to check the correctness and accuracy of the numerical technique used. The numerical values where round to five decimal places and they show a good and exact agreement with the exact result via Eq. (\ref{omega}). As it was mentioned, these oscillations will be damped leading to stable perturbations for $2 \Lambda n (1+n) > m^2$. However, for $2 \Lambda n (1+n) < m^2$, the oscillations would lead to unstable modes.  




\begin{table}[ht]
\caption{QNFs 
for massless  scalar fields with $\Lambda=1, m=3$ in the background of tree-dimensional extreme dilaton black holes.
}
\label{Table1}\centering
\begin{tabular}{ | c | c | c | }
\hline
$n$ & $\omega_{AIM}$ & $\omega$  \\\hline
$0$ & $18.00000i$ & $18i$  \\
$1$ & $3.33333i$ & $10/3i$ \\
$2$ & $-1.20000i$ & $-6/5i$   \\
$3$ & $-4.28571i$ & $-30/7i$ \\
$4$ & $-6.88889i$ & $-62/9i$   \\\hline
\end{tabular}
\end{table}

\newpage


\section{Conclusion}
\label{conclusion}

In this work we considered the propagation of massless scalar fields in the background of three-dimensional extremal dilaton black holes. We obtained QNM frequencies analytically, and we showed that the QNMs are overdamped or purely imaginary leading to stable perturbations for $2 \Lambda n (1+n) > m^2$. For $2 \Lambda n (1+n) < m^2$ the oscillations would lead to unstable modes. The decay rate increases when the cosmological constant increases. For the fundamental mode $n=0$, the decay rate depends only on the angular number of the test field: however in this case the modes are  unstable. Also, we used the improved AIM in order to determine  the QNFs numericaly, and we showed that there is a good agreement between the numerical and the analytical solutions. \\

There are number of works that are interesting to do in the future related to this work presented here: Onozawa et.al \cite{onozawa2} showed that QNM's for the extreme Reissner-Nordstr\"om black holes for the spin 1, 3/2, 2 are the same. It  would be interesting to see if that is the case for the extreme dilaton black hole. It would also be interesting to analyze the superradiant instability \cite{Brito:2015oca} of this  extremal black hole for charged massive scalar field, as well as, the quasinormal modes:  we leave those  for a future work, see \cite{Gonzalez:2021vwp}. \\



\end{document}